# Pressure induced quantum phase transitions

G. A. Gehring

Department of Physics and Astronomy

University of Sheffield

Sheffield S3 7RH UK

**Abstract**

A quantum critical point is approached by applying pressure in a number of ferromagnetic and antiferromagnetic metals. The observed dependence of $T_c$ on pressure necessarily means that the magnetic energy is coupled to the lattice. A first order phase transition occurs if this coupling exceeds a critical value: this is inevitable if $\frac{\partial T_c}{\partial p}$ diverges as $T_c$ approaches zero. It is argued that this is the cause of the first order transition that is observed in many systems. Landau theory is used to obtain the phase diagram and also to predict the regions where metastable phases occur that agree well with experiments done on MnSi and other materials. The theory can be used to obtain very approximate values for the temperature and pressure at the tricritical point in terms of measured quantities. The values of the tricritical temperature for various materials obtained from Landau theory are too low but it is shown that the predicted values will rise if the effects of fluctuations are included.

PACS numbers: 05.70.Fh, 64.70.Tg, 75.20.en

The study of metallic systems when the transition temperature is driven to zero is exciting much interest because the character of a phase transition changes due to fluctuations being caused by quantum effects rather than the thermal fluctuations that dominate normal phase transitions. This is known as a quantum critical point (QCP) and occurs when two ground states are degenerate. In a metallic compound Fermi liquid theory breaks down at a QCP transition between a magnetic and nonmagnetic state because the strong quantum fluctuations couple to the electrons; a superconducting phase often occurs. Disorder is a relevant variable so experimental probes are sought that will tune a transition temperature to zero while retaining the crystal homogeneity. A very attractive method is to use hydrostatic pressure because this increases the band width in a metallic magnet and hence reduces the magnetic energy. A number of materials, MnSi [1,2,3], $UGe_2$ [4], $CeIn_3$ [5], $ZrZn_2$ [6,7] and $URu_2Si_2$ [8], have been found that are close enough to a critical density so that the application of an accessible pressure reduces the transition temperature to zero. One of the points that is becoming clear is that in these materials a first order transition occurs before the QCP is reached [3], also the distortion energy itself is very significant and should be included in the theory [2].

These materials are special for two reasons: first because the pressure dependence of the transition temperature, is measured directly and secondly because it becomes large, and in many cases appears to diverge at the QCP. In this letter we show that Landau theory [9] gives a good account of the first order transition seen in compressible magnets near the QCP . The calculation is presented in three parts; first we derive an expression for the pressure and temperature at the point where the transition become first order, the tricritical point. Metastable phases occur when the transition is first order so that phase separation occurs; the phase diagram where phase separation may be observed is calculated. Finally we discuss how the relevant Landau parameters may be estimated and so use the theory to estimate values of the pressures and temperatures at the tricritical points due to this mechanism. We find that the prediction of the parameters of the tricritical point is lower than experiment but the origin of this is understood.

We note that Belitz et al [10] have shown quantum fluctuations of the electron gas always induce a first order phase transition in ferromagnets approaching a QCP. It has also been shown that there is a first order transition if the electronic density of states has a minimum at the Fermi energy [11,12]. In both of these cases the Landau parameter $B$ falls to zero as the temperature is reduced.

The change in the critical properties due to effects arising from a compressible lattice and the effects of pressure on phase transitions have been investigated extensively over many years. In magnetothermodynamics (MTM) [13] the magnetic Hamiltonian is assumed to depend linearly on strain and it is found that the results for a compressible lattice at temperature $T$ and pressure $p$ may be mapped on to an incompressible lattice at temperature $T^*$. A first order transition is inevitable if the magnetic specific heat diverges on an incompressible lattice or if $\frac{\partial T_c}{\partial p}$ is large enough. This conclusion was confirmed when lattice vibrations were included in a renormalisation group treatment of the Ising model [14]. Unfortunately the method used in MTM is inapplicable to materials approaching a QCP because of the assumptions that there is a single magnetic Hamiltonian that is renormalized by coupling to the lattice that implies that the entropy per spin, $S$, of the disordered phase is $k_B \text{Log}(2S+1)$ independent of the value of $T_c$. An essential feature of a QCP is that the Hamiltonian has *two* competing components each of which can lead to a ground state that satisfies the third law of thermodynamics and that the effect of pressure is to change the relative energies of these two quantum ground states. This necessarily means that the magnitude of both the order parameter and the entropy difference between the ordered and non-ordered state should vanish as the QCP is approached. MTM theory violates these conditions because so long as the transition temperature is finite the saturation magnetisation is independent of the transition temperature as is the entropy difference between the fully ordered and disordered states.

In Landau theory the nature of the phase transition is unchanged when the transition temperature depends on pressure until a critical value of $\frac{\partial T_c}{\partial p}$ is reached when the transition becomes first order [9]. It is almost always true that if a phase transition is first order within Landau theory it will be first order when a full renormalisation group theory is undertaken (an exception to this occurs for the three state Potts model in two dimensions). This is because the effects of the renormalisation close to the critical point are to accelerate the transition to a first order transition. Later we review the evidence that at least one material, ZrZn$_2$, is surprisingly well described by Landau theory in the second order region. The tricritical point is the end point of the second order line and so the theory is justified as this point is approached.

We now address the validity of the Landau theory where the transition is first order. The line of first order transitions is a continuation of the second order line indicating that the nature of the ordered phase is the same when the transition to the ordered phase is first or second order. The magnitude of the order parameter in one example, MnSi, measured at low temperatures changes smoothly with pressure through the tricritical point [15]. Hence we can assume that the low temperature phase may be obtained in terms of an expansion in terms of the order parameter about the high temperature phase.

The Gibbs free energy is a minimum at a given temperature and pressure and in Landau theory is expressed in terms of two variational parameters, the compressive strain, $\varepsilon = -\frac{\delta V}{V}$, and the magnetisation, $M$, both of which are eliminated by minimisation. In this expression the transition temperature is given in terms of the strain, $T_c(\varepsilon)$ because this is what would be given in a microscopic theory [9]. The Landau expansion for the Gibbs free energy is given below,

$$F(M,\varepsilon:p,T) = \frac{A}{2}(T-T_c(\varepsilon))M^2 + \frac{B}{4}M^4 + \frac{1}{2}K\varepsilon^2 - p\varepsilon. \tag{1}$$

In this equation $p$ is the pressure and $K$ the bulk modulus; the parameters $A$ and $B$ have been assumed to be independent of pressure and temperature; a justification for this is given later. We chose to write Eq. (1) in terms of a scalar magnetisation for convenience – all the results would follow for an isotropic magnet when $M^2$ was replaced in Eq. (1) by $\mathbf{M}^2$ and $M^4$ by $(\mathbf{M}^2)^2$ and are also valid for an antiferromagnet if $M$ represents a sublattice magnetisation and $T_c(\varepsilon)$ represents the Néel temperature.

The expression for the tricritical point is the same whichever is eliminated first $M$ or $\varepsilon$. The derivation where $M$ is eliminated first leads naturally to a description of the phase diagram so it is presented here. The free energy has a turning point where,

$$\frac{\partial F(M,\varepsilon,p,T)}{\partial M} = 0 = M\left[A(T-T_c(\varepsilon)) + BM^2\right]. \tag{2}$$

This has two solutions, the nonmagnetic phase $M=0$ which is a solution at all temperatures and the low temperature solutions $M = \pm\sqrt{\frac{A(T_c(\varepsilon)-T)}{B}}$ that are allowed provided that $T < T_c(\varepsilon)$.

In the nonmagnetic phase the equilibrium value of the strain is given by $\varepsilon = \frac{p}{K}$ and the Gibbs free energy is found to be, $F_0(p,T) = -\frac{p^2}{2K}$ as this is independent of temperature it is written as $F_0(p)$.

The free energy of the magnetic phase, $F_M$ (valid in the range $T < T_c(\varepsilon)$), is a function of strain which is found by minimisation,

$$F_M(\varepsilon:p,T) = -\frac{A^2}{4B}(T_c(\varepsilon)-T)^2 + \frac{1}{2}K\varepsilon^2 - p\varepsilon. \tag{3}$$

$$\varepsilon - \frac{p}{K} = \frac{A^2}{2BK}(T_c(\varepsilon)-T)\frac{\partial T_c(\varepsilon)}{\partial \varepsilon} = \delta(\varepsilon) \tag{4}$$

In this expression δ(ε) is a contribution to the strain due to exchangestriction; it is negative because the transition temperature falls as the strain increases, $\frac{\partial T_c}{\partial \varepsilon} < 0$; strain inhibits magnetisation and so magnetisation reduces strain. The Gibbs free energy in the magnetic phase contains three terms, the free energy in the nonmagnetic phase, $F_0(p)$, the magnetic term which is negative definite below $T_c$ and an extra contribution arising from the strain dependence of the order parameter, the value of δ(ε) in this expression is evaluated at equilibrium from Eq. (4).

$$F_{mag}(p,T) = F_0(p) - \frac{A^2}{4B}(T_c(\varepsilon) - T)^2 + \frac{K\delta(\varepsilon)^2}{2}. \tag{5}$$

This leads to the central result of this Letter that the stable phase will be magnetic provided *both* $T < T_c(\varepsilon)$ and $F_{mag}(p,T) < F_0(p)$.

The condition for a first order transition is that the Gibbs free energy is the same in two phases, $F_{mag}(p,T) = F_0(p)$. Combining the results of equations 4 and 5 we obtain the following expression for the free energy,

$$F_{mag}(p,T) = F_0(p) - \frac{A^2}{4B}(T_c(\varepsilon) - T)^2 \left[1 - \frac{A^2}{2KB}\left(\frac{\partial T_c}{\partial \varepsilon}\right)^2_{\varepsilon_1}\right] \tag{6}$$

The second term vanishes both on the second order line, $T = T_c(\varepsilon)$, and also for a special value of the strain, $\varepsilon = \varepsilon_1$.

$$\frac{A^2}{2KB}\left(\frac{\partial T_c}{\partial \varepsilon}\right)^2_{\varepsilon_1} = 1 \tag{7}$$

The tricritical point occurs where Eq. (7) is valid for the same conditions as are required for a second order transition namely, $T = T_c(\varepsilon)$, $M=0$ and $\varepsilon = \frac{p}{K}$. This means that the tricritical point occurs for a pressure, $p_t$ such that $\frac{A^2 K}{2B}\left(\frac{\partial T_c}{\partial p}\right)^2_{p=p_t} = 1$. On dimensional grounds we can write $\left(\frac{\partial T_c}{\partial p}\right)^2_{p=p_t} = \frac{T_c^2}{p_0^2} F\left(p/p_0\right)$. This enables us to write the condition for the tricritical point as $\Theta F\left(p_t/p_0\right) = 1$. The parameter $\Theta \left(= \frac{A^2 T_c^2 K}{2 B p_0^2}\right)$ is the ratio of the magnetic energy at zero temperature and pressure ($-\frac{A^2 T_c^2}{4B}$) to the strain energy at a pressure $p_0$ ($-\frac{p_0^2}{2K}$); a first order transition occurs more readily if $\Theta$ is large. In many cases $T_c(p) = T_0 \left(1 - p/p_0\right)^\zeta$ where $T_0$ is the transition temperature in the absence of applied pressure and $p_0$ is a critical pressure where the exponent $\zeta$ is less than unity; ¾ is predicted by theory [1], hence $\frac{\partial T_c}{\partial p}$ diverges as the QCP is approached so a first order transition is inevitable. In this case $F\left(p/p_0\right) = \zeta^2 \left(1 - p/p_0\right)^{2\zeta - 2}$ so the tricritical pressure $p_t$ is given by $\left(1 - p_t/p_0\right) = \left(\Theta \zeta^2\right)^{\frac{1}{2(1-\zeta)}}$; the temperature at the tricritical point is found from $T_t = T_0 \left(1 - p_t/p_0\right)^\zeta$.

The theory is extended to calculate the phase diagram for temperatures below $T_t$. The condition given by Eq. (6) predicts that the first order transition occurs at the strain $\varepsilon_1$ for all temperatures below $T_t$ as shown in Fig. 1a. This is what would be observed if the experiment was performed at constant strain (for example if the sample were embedded in solid helium). The value of $\varepsilon_1$ is found immediately from eq. 7 by $K\varepsilon_1/p_0 = 1 - \left(\Theta \zeta^2\right)^{\frac{1}{2(1-\zeta)}}$ if the transition temperature follows the relation assumed earlier, $T_t = T_0 \left(1 - p_t/p_0\right)^\zeta$.

The theory has shown us that if the strain in the magnetic phase is given by $\varepsilon = \varepsilon_1$ then $F_{mag}(p,T)$

$=F_0(p)$. The *two* distinct macroscopic states are defined at the same pressure and temperature

where $T < T_t$. . This defines the line of first order phase transitions which is calculated now.

The magnetism in the magnetic phase has strain $\varepsilon_1$ and is given by,

$$M_t(T) = \pm\sqrt{\frac{A(T_c(\varepsilon_1)-T)}{B}} = \pm\sqrt{\frac{A(T_t-T)}{B}} \quad . \tag{8}$$

The pressure of the magnetic phase is found from Eq. (4) where the strain is set to its critical value, $\varepsilon = \varepsilon_1$.

$$p_1(T) = p_t - \frac{A^2}{2B}(T_t-T)\frac{\partial T_c(\varepsilon)}{\partial \varepsilon}\bigg)_{\varepsilon=\varepsilon_1} = p_t - (T_t-T)\left(\frac{\partial T_c}{\partial p}\right)_{p_t}^{-1} \tag{9}$$

The free energy of this magnetic phase is equal to that of a phase with zero magnetisation also at pressure $p_1(T)$.

This relation shows that the line of first order transitions is defined by $p_1(T)$ which varies linearly with temperature with a slope given by the tangent to the second order phase line at the tricritical point. The low temperature limit of this phase line is given by $p_1(0) = p_t - \frac{A^2 T_t}{2B}\frac{\partial T_c(\varepsilon)}{\partial \varepsilon}\bigg)_{\varepsilon=\varepsilon_1}$.

It is straightforward to understand physically what is happening here. The transition temperature is *falling* with increased strain and so the magnetic energy will be increased if the strain is reduced. This means that the state with finite magnetisation is *less* strained (expanded) relative to the nonmagnetic state. At the first order transition we have a contracted phase with zero magnetisation and a magnetic, expanded state. We note that the transition temperature is a very strong function of strain near to the QCP, $T = T_0(1-\varepsilon/\varepsilon_0)^\zeta$. As the system becomes magnetic near to the QCP the strain, $\varepsilon$, is reduced, the transition temperature rises and so the magnetism increases further. This is the positive feedback that fuels the first order transition and explains why there are no allowed states with magnetisations between $M=0$ and $M_t(T) = \pm\sqrt{\frac{A(T_t-T)}{B}}$ on the first order line.

We now consider the regions where metastable states can exist. The nonmagnetic state is an allowed solution for all strains $\varepsilon \geq \varepsilon_1$ and the pressure in this state is given by $p = K\varepsilon$. In the range $p_t < p < p_1(T)$ the free energy in the nonmagnetic state is higher than that in the magnetic state because $\frac{A^2}{2KB}\left(\frac{\partial T_c}{\partial \varepsilon}\right)^2_{\varepsilon_1} < 1$ as is seen from Eq. 6. Hence in the region defined by $p_t < p < p_1(T)$ the nonmagnetic state is metastable and the magnetic state is stable.

We now consider the stability of metastable states in the range $p_t < p < p_1(T)$. In this region there are two local minima in the free energy separated by a region of higher, (locally a maximum) energy. One region has $M=0$ and a higher value of the strain, the other has a finite magnetisation and a lower value of the strain. In general any first order transition that involves a change in the lattice shows strong hysteresis and also a mixed phase region. The most well known example is that of a martensitic phase transition which involves a change of symmetry – a better example is the α to γ transition of metallic Ce (also a volume change) – which also shows metastable states and hysteresis [16]. This fact reflects the long range interactions between regions of the sample with different strain – as one part strains it will cause an additional stress on other parts of the sample. The magnitude of the local maximum in the free energy can be estimated from the different elastic energies of the two phases. On the first order line the strain in the magnetic phase is $\varepsilon_1$ and that in the nonmagnetic state is given by $\varepsilon = p_1(T)/K$. The difference in the elastic energies is given by $\frac{(T_t - T)^2}{2K}\left(\frac{\partial T_c}{\partial p}\right)^{-2}_{p_t}$ - this difference vanishes at the tricritical point but is increased at low temperatures which is just where the thermal energy available to excite the system over the barrier is becoming unavailable. Hence hysteresis will be seen most strongly at the lowest temperatures. As the pressure is reduced from $p_1(T)$ towards $p_t$ the difference between the free energies of the metastable, nonmagnetic, state and the stable, magnetic, state is increasing. The magnitude of the local maximum in the free energy is also falling through this region so that less of the sample will be in the nonmagnetic state. The nonmagnetic state becomes unstable at $p_t$. The value of the magnetism in the magnetic component will rise slowly from its value on the first order line as the pressure is reduced towards $p_t$ because the strain is reducing and hence $T_c(\varepsilon)$ is rising. The magnetisation in the nonmagnetic state remains at zero. The phase diagram is shown in Fig 1b. This is what is seen in MnSi [1, 3, 15] reports also are given of mixed phases in URu$_2$Si$_2$ [8], UGe$_2$[4], ZrZn$_2$ [7], CeIn$_3$ [5] and many other materials [3].

For pressures above $p_1(T)$ the nonmagnetic state has the lowest free energy but there will be a region where the magnetic state is metastable. This is found from the value of the maximum pressure for which Eq. 4 may be satisfied. This occurs at a strain $\varepsilon^*$ obtained from Eq 4 using $\left.\frac{\partial p}{\partial \varepsilon}\right)_{\varepsilon=\varepsilon^*} = 0$. This region will be relatively small because the value of $T_c(\varepsilon)$ is falling as the strain increases above $\varepsilon_1$. This phase may be responsible for the region above the first order transition where anomalous inelastic scattering is observed [17].

We note that the phase diagram shown in Fig 1b differs from that obtained from a Landau theory with a negative fourth order term, ($B=-B'$), $F_- = \frac{A(T-T_c)M^2}{2} - \frac{B'M^4}{4} + \frac{CM^6}{6}$. In this case the first order transition temperature is at $T_1 = T_c\left(1 + \frac{3B'^2}{16AC}\right)$ and hysteresis may occur between $T_L$ and $T_U$ where $T_L = T_c$ and $T_U = T_c\left(1 + \frac{B'^2}{4AC}\right)$. We note that all of these temperatures are greater or equal to $T_c$ which is the temperature at which the second order transition would have occurred had the fourth order term been positive. There are two reasons why the fourth order term may fall to zero at low temperatures. This will occur if there is a local minimum in the density of states at the Fermi energy in the paramagnetic phase. The fourth order term is driven to zero by quantum fluctuations in a metallic system and the phase line moves to higher pressures as is shown in Fig 2 ($G=0$) in [10].

The third law of thermodynamics imposes a constraint on the bare parameter $B$. Both the saturation magnetisation and the change in entropy on entering the ordered phase, $M(T=0) = \sqrt{AT_c(\varepsilon)/B}$ and

$$\Delta S_{mag}(0) = -\left.\frac{\partial F_{mag}}{\partial T}\right)_{T=0} = -\frac{A^2 T_c(\varepsilon)}{2B}$$

must tend to zero as $T_c \to 0$ because the ordered phase must evolve continuously into a coherent quantum state as required by the third law of thermodynamics. This shows that $B$ (defined at constant ε in the ordered phase) is either a constant or depends weakly on $T_c(\varepsilon)$.

The most direct evidence of Landau behaviour in weak metallic ferromagnets, at zero pressure, comes from studies of ZrZn$_2$ as a function of temperature and magnetic field [18,19]. Arrott plots of μ$_0$H/ M against $M^2$ gives a straight line if Landau theory is valid with slope $B$ and intercept $A(T-T_c)$. Parallel straight lines were found for different temperatures which showed that the Landau parameter $B$ was approximately independent of temperature and the specific heat jump was in agreement with that predicted from Landau theory with these values of the parameters [19]. This is important as the theory described here is applicable to a material that is better described using Landau theory than by Slater theory. We have assumed that the Landau parameter $A$, which is inversely proportional to the Curie Weiss susceptibility, is independent of pressure. This has been verified by susceptibility measurements as a function of temperature and pressure in the paramagnetic phase of MnSi [1] and ZrZn$_2$ [20].

The value of Θ may be estimated from experiment; $T_c$ and $p_0$ are measured directly, the compressibility $K$ is known, or may be estimated, so it remains to estimate $\frac{A^2}{B}$. The most direct method is to use fit the specific heat discontinuity at the transition at ambient pressure, $\Delta C_H(T_c) = \dfrac{T_c A^2}{2B}$ this is valid for both antiferromagnets as well as ferromagnets. The two parameters, *A* and *B* may be determined separately for ferromagnets using an Arrott plot, this was done for ZrZn$_2$ [19]. Alternatively the parameter A is found from the susceptibility in the paramagnetic phase: $A(T - T_c) = \mu_0 \chi_{C-W}^{-1} = \dfrac{3k_B(T-T_c)}{Nm_{eff}^2}$ where *N* is the number of magnetic ions per unit cell, $m_{eff}$ is the effective moment and $k_B$ is Boltzmann's constant. The coefficient *B* may be estimated from the saturation magnetisation, $Nm_0 = \sqrt{\dfrac{AT_c}{B}}$, where $m_0$ is the magnetic moment per unit cell, this seriously underestimates *B* because the magnetisation does not follow the Landau theory prediction down to low temperatures. Combining these expressions we obtain a useful relation $\dfrac{A^2}{2B} = \dfrac{3k_B N}{2T_c}\left(\dfrac{m_0}{m_{eff}}\right)^2$.

In the case of ZrZn$_2$ the Landau parameters have been determined from experiment directly [19] $A=8.2\times 10^{-6}$ [T/KAm$^{-1}$] and $B=1.1\times 10^{-13}$ T[Am$^{-1}$]$^{-3}$. *K* is taken as 135GP, the value for MnSi, $p_0$ ~1.6GPa and $T_c = 25$K; these values lead to an estimated value of $\Theta$~ 0.02 or ~0.04 for ZrZn$_2$ if the change of *B* with lattice constant is included [21]. Values of $\Theta$ for other materials MnSi, CeIn$_3$ and UGe$_2$ were evaluated using magnetic data also taking K =135GPa. Data for MnSi are $m_0/m_{eff}$ ~ 0.4/1.4, $N=4.25\times 10^{28}$m$^{-3}$, $p_0$ ~1.6GPa and $T_c = 28$K [1] which leads to a value of $\Theta$ ~0.3 ; using $m_0/m_{eff}$ ~ 0.5/0.7, $p_0$ ~2.65GPa, $N=1.6\times 10^{28}$m$^{-3}$, with $T_c = 10$K [5] we find $\Theta$ ~0.02 for CeIn$_3$. The magnetisation in UGe$_2$ has two first order transitions [4] so is more difficult to fit to this theory, but using $m_0/m_{eff}$ ~ 0.5/0.7, $p_0$ ~1.6GPa, $N=1.6\times 10^{28}$m$^{-3}$ and $T_c = 53$K we find $\Theta$ ~0.05. The values of $T_t$ and $p_t$ obtained from $\Theta$ are extremely approximate because the low temperature magnetisation was used to obtain *B* and also because the values depend strongly on the value of $\zeta$ which is known very approximately. The values obtained are consistently low.

The most severe approximation in Landau theory is that there were no magnetic contributions to the Gibbs free energy in the high temperature phase. If we write the free energy in the high temperature phase as $F(T > T_c(\varepsilon)) = -\frac{p^2}{2K} - F_+ \left(\frac{T-T_c}{T_c}\right)^{2-\alpha}$ then the condition given by Eq. 6 is changed to $\frac{A^2}{2KB}\left(\frac{\partial T_c}{\partial \varepsilon}\right)^2_{\varepsilon_1} = 1 - F_+ \frac{4B}{A^2}\left(\frac{T-T_c}{T_c}\right)^{-\alpha}$ which will reduce the value of $\left(\frac{\partial T_c}{\partial \varepsilon}\right)^2_{\varepsilon_1}$ required for a first order transition.

Another approximation is that the compressibility was assumed to be independent of the pressure – at these high pressures anharmonic effects may becoming important especially as the electronic structure is changing so drastically.

A QCP may be approached by alloying and expected QCP behaviour was observed in $Zr_{1-x}Nb_xZn_2$ [22] and $Pd_{1-x}Ni_x$ [23]. However first order transitions and phase separation were reported in a number of other alloyed systems [3]. It needs to be determined whether the relevant variable is quenched, as is the case for atomic disorder, or is annealed as is the case if the important variable is the average electron density [24].

The most important successes of this theory are that it predicts universality of a first order transition, the form of the phase diagram and phase separation for both ferromagnets and antiferromagnets provided that $\frac{\partial T_c}{\partial p}$ diverges as the critical point is approached. If there are other effects [10,11,12] that also drive the material to a first order transition then the effects of the compressible lattice will also accelerate the trend to a tricritical point. The effects of the compressible lattice should be incorporated into these theories.

It is becoming clear that phase separation is a near universal phenomenon near to QCP's [3]. Theory predicts this for measurements made as a function of pressure and found that the critical pressure for the onset of the phase separation depends linearly on temperature in agreement with experiment [15].

In summary the important points of the Letter are that it demonstrates that simple Landau theory, without including fluctuations, predicts the following results. A first order transition will occur before a quantum critical point that is accessed by the application of pressure and that the coordinates of the tricritical point made be estimated from the observed dependence of the transition temperature, and other parameters of the magnetic phase. This will occur for any type of magnetic order, not just for ferromagnets. Different phase diagrams will occur if the temperature is lowered at constant strain instead of constant pressure. The region where metastable phases can occur is also calculable from the observed dependence of $T_c$ on pressure.

Grateful acknowledgments are due to SM Hayden for helpful correspondence on the transition in $ZrZn_2$ and many discussions with SS Saxena and GG Lonzarich who also made helpful comments on the draft manuscript.

**Figure Captions**

Fig. 1a The phase diagram as a function of strain. The first order line (shown wider) occurs at strain $\varepsilon_1$ for all temperatures.

Fig. 1b The phase diagram as a function of pressure. The first order line (shown wider) is given by $p_1(T)$. There is a region between $p_t$ and $p_1(T)$ where the magnetic phase is stable and the nonmagnetic phase is metastable.

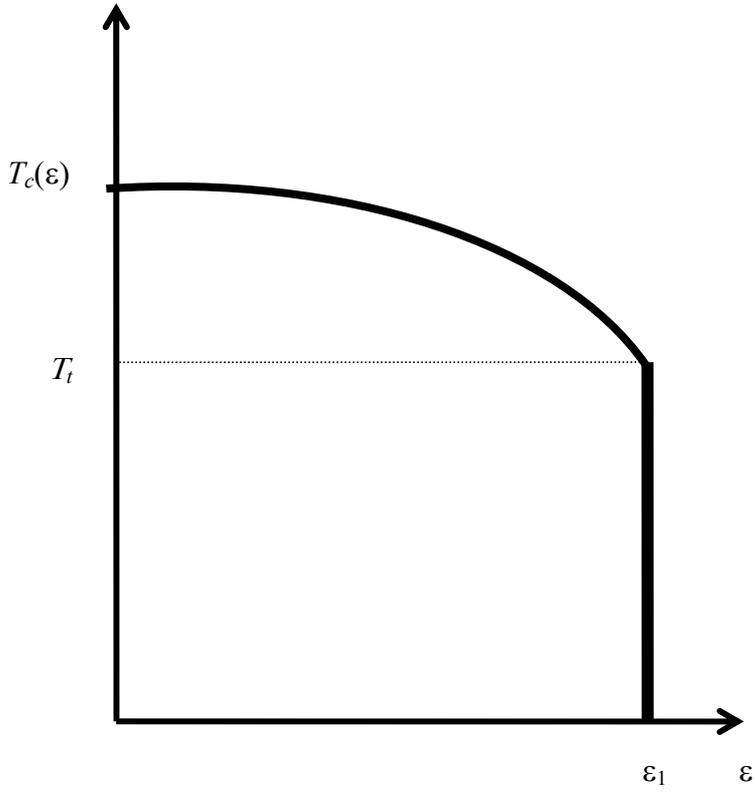

Figure 1a

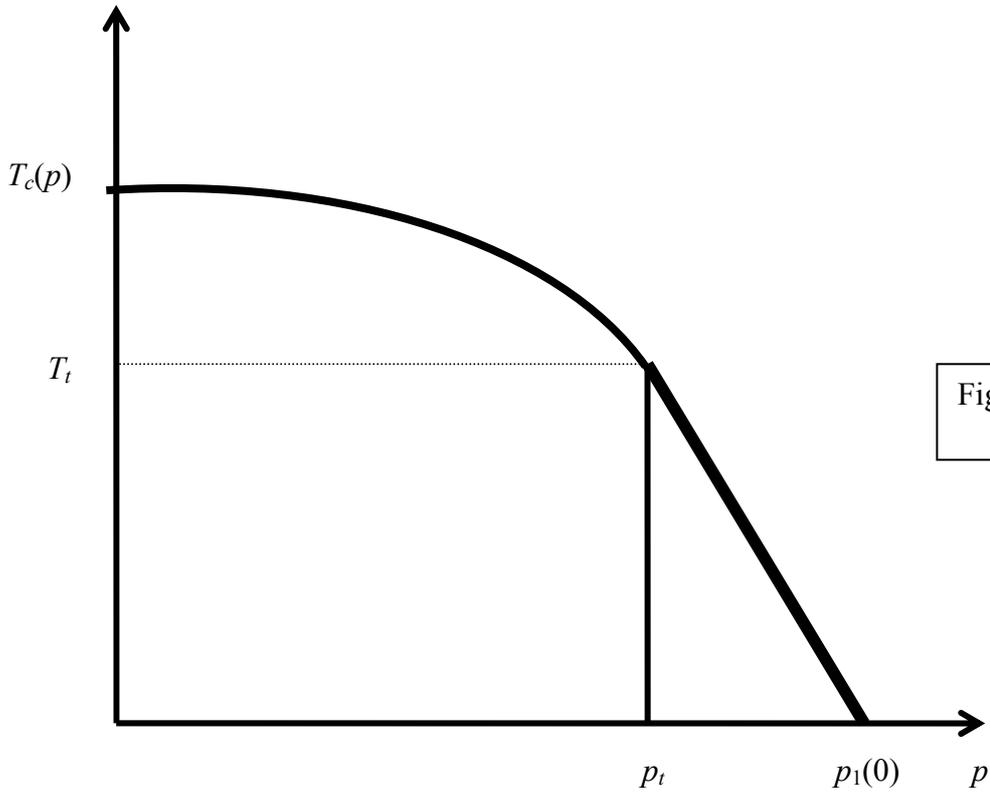

Figure 1b